\documentclass[pra,showpacs]{revtex4}

\usepackage{bm}
\usepackage{graphicx}
\usepackage{amssymb}
\usepackage{amsmath}

\begin{document}

\title{Multipartite entanglement, quantum-error-correcting codes, and entangling power of quantum evolutions}

\author{A. J. Scott}
\email{ascott@phys.unm.edu} \affiliation{Department of Physics
and Astronomy, University of New Mexico, Albuquerque, NM 87131-1156,
USA}

\begin{abstract}
We investigate the average bipartite entanglement, over all possible divisions of a 
multipartite system, as a useful measure of multipartite entanglement. We expose 
a connection between such measures and quantum-error-correcting codes by deriving 
a formula relating the weight distribution of the code to the average entanglement 
of encoded states. Multipartite entangling power of quantum evolutions is also 
investigated.
\end{abstract}

\maketitle

\def\w{\omega}
\def\wb{\overline{\omega}}
\def\tr{\operatorname{tr}}
\def\Tr{\operatorname{Tr}}
\def\wt{\operatorname{wt}}
\def\ket#1{|#1\rangle}
\def\bra#1{\langle#1|}
\def\inner#1#2{\langle #1 | #2 \rangle}

\section{Introduction}
\label{sec1}

The phenomenon of entanglement \cite{horodecki,wootters,horodecki2} is a remarkable feature of 
quantum physics that has been identified as a key ingredient in many areas of quantum 
information theory including quantum key distribution \cite{ekert}, superdense coding \cite{bennett2} 
and teleportation \cite{bennett3}. However the general problem of how to quantify \cite{horodecki} the 
level of entanglement in an arbitrary multipartite system remains unresolved. There 
has been some progress towards a solution \cite{dur,bennett4,eisert,verstraete,verstraete2,miyake,miyake2,jaeger,jaeger2,jaeger3,bravyi}, 
but the task at hand is generally considered a difficult one and may never be completed. 
We are thus lead to consider simple computable measures of entanglement \cite{wong,meyer} 
that although cannot fully characterize the multipartite nature of the correlations, may 
nevertheless still provide a useful gauge of their levels. 

In this article we investigate the average bipartite entanglement, over all possible divisions of a 
multipartite system, as a useful measure of multipartite entanglement. Such measures might be 
considered the least sophisticated of choices; however, their simplicity allows theoretical calculations to be 
exercised with ease. We will restrict our study 
to pure-state entanglement where the subsystem linear entropy is a clear choice for the bipartite measure. 
It was recently shown by Brennen \cite{brennen} that an entanglement measure proposed by Meyer and Wallach 
\cite{meyer} is of the above described form, and hence, the multipartite entanglement measures 
considered in this paper may be viewed as generalizations of the Meyer-Wallach measure. Our measures 
may also be viewed as variations of those considered by Pope and Milburn \cite{pope} where instead 
the minimum bipartite entanglement was considered. 

We show that the average bipartite entanglement elects self-dual quantum-error-correcting 
codes to the status of maximally entangled states. The connection between entanglement 
and quantum-error-correcting codes has been highlighted elsewhere (e.g. \cite{preskill}); 
however we make this relationship explicit by expressing the average 
entanglement of encoded states in terms of the weight distribution of the code. We also 
investigate the multipartite entangling power of quantum evolutions. A simple extension 
of the work of Zanardi {\it et al.} \cite{zanardi} allows the derivation of an explicit 
formula. Such formulae are relevant to current studies in the entangling capabilities 
of chaotic systems \cite{sakagami,tanaka,furuya,angelo,miller,lakshminarayan,bandyopadhyay,tanaka2,fujisaki,lahiri,lakshminarayan2,bettelli,scott,bandyopadhyay2,jacquod,rossini}.
An example treated in this article is the quantum kicked rotor.
 
The paper is organized as follows. In the next section we introduce the Meyer-Wallach 
entanglement measure and its generalizations. The connection between these 
measures and quantum-error-correcting codes is discussed in Section \ref{sec3}.
This relationship is further strengthened in Section \ref{sec4} where we derive 
a formula for the average entanglement over a subspace. In Section \ref{sec5} 
we derive a formula for the multipartite entangling power of an arbitrary 
unitary. Finally in Section \ref{sec6} we conclude by applying our results 
to the quantum kicked rotor.

\section{A class of multipartite entanglement measures}
\label{sec2}

It is generally accepted that when a bipartite quantum system is in an overall 
pure state, there is an essentially unique resource-based measure of entanglement
between the two subsystems.  This measure is given by the von Neumann
entropy of the marginal density operators \cite{bennett,popescu}. To ease 
theoretical calculations, one often replaces the von Neumann entropy with its 
linearized version, the linear entropy. For a bipartite system in an overall pure 
state $\ket{\psi}\in\mathbb{C}^{D_A}\otimes\mathbb{C}^{D_B}$, 
the {\it subsystem linear entropy} is defined as
\begin{equation}
S_L(\psi)\equiv\eta\left(1-\tr{\rho_A}^2\right) \qquad \rho_A=\tr_B\ket{\psi}\bra{\psi}
\end{equation}
where the normalization factor, $\eta=D/(D-1)$ with $D=\min(D_A,D_B)$, is chosen 
such that $0\leq S_L\leq 1$. The state is separable if and only if $S_L=0$, and maximally 
entangled when $S_L=1$ .

In general, as the number of subsystems increases, an exponential number of
independent measures is needed to quantify fully the amount
entanglement in a multipartite system.  Consequently, the 
following entanglement measures cannot be thought of as unique.
Different measures will capture different aspects of multipartite
entanglement.

The Meyer-Wallach measure \cite{meyer}, $Q(\psi)$, which can only be applied to
multi-qubit pure states $\ket{\psi}\in(\mathbb{C}^2)^{\otimes n}$, is defined as follows.  For each
$j=1,\dots,n$ and $b\in\{0,1\}$, we define the linear map
$\imath_j(b):(\mathbb{C}^2)^{\otimes n}\rightarrow(\mathbb{C}^2)^{\otimes n-1}$ 
through its action on the product basis,
\begin{equation}
\imath_j(b)\ket{x_1}\otimes\dots\otimes\ket{x_n}=
\delta_{bx_j}\ket{x_1}\otimes\dots\otimes\ket{x_{j-1}}\otimes
\ket{x_{j+1}}\otimes\dots\otimes\ket{x_n}\;
\end{equation}
where $x_i\in\{0,1\}$. The Meyer-Wallach entanglement measure is then
\begin{equation}
Q(\psi)\equiv
\frac{4}{n}\sum_{j=1}^n
D\big(\imath_j(0)\ket{\psi},\imath_j(1)\ket{\psi}\big)
\end{equation}
where
\begin{equation}
D\big(\ket{\psi},\ket{\phi}\big)=\inner{\psi}{\psi}\inner{\phi}{\phi}-|\inner{\psi}{\phi}|^2.
\end{equation}
Meyer and Wallach showed that $Q$ is invariant under local
unitary transformations and that $0\leq Q\leq 1$, with $Q(\psi)=0$ if
and only if $\ket\psi$ is a product state. 

Recently, it was shown by Brennen \cite{brennen} that $Q$ is simply the average 
subsystem linear entropy of the constituent qubits:
\begin{equation}
Q(\psi)=2\Bigg(1-\frac{1}{n}\sum_{k=1}^n\tr{\rho_k}^2\Bigg) \label{brennenformula}
\end{equation}
where $\rho_k$ is the density operator for the $k$-th qubit after tracing out the rest. This 
simplification is easily understood \cite{caves} by first showing that 
$D(\imath_j(0)\ket{\psi},\imath_j(1)\ket{\psi})$ is unchanged by a local unitary applied to the 
$j$-th qubit (a fact already proven by Meyer and Wallach), and hence, invariant under a change in the 
qubit's fiducial basis. Consequently, a judicious choice of the Schmidt basis gives 
$D(\imath_j(0)\ket{\psi},\imath_j(1)\ket{\psi})=\lambda_j^1\lambda_j^2=(1-\tr{\rho_j}^2)/2$,  
where $\lambda_j^1$ and $\lambda_j^2$ are the Schmidt coefficients in the decomposition between the 
$j$-th qubit and the remainder of the system.

Brennen's simplification immediately allows the generalization of $Q$ 
to multi-qudit states $\ket{\psi}\in(\mathbb{C}^D)^{\otimes n}$, and by considering 
all other possible bipartite divisions, we can now define a class of related 
multipartite entanglement measures in the obvious manner:
\begin{equation}
Q_m(\psi)\equiv\frac{D^m}{D^m-1}\Bigg(1-\frac{m!(n-m)!}{n!}\sum_{|S|=m}\tr{\rho_{S}}^2\Bigg) 
\qquad m=1,\dots,\lfloor n/2 \rfloor
\end{equation}
where $S\subset\{1,\dots,n\}$ and $\rho_S = \tr_{S'}\ket{\psi}\bra{\psi}$ 
is the density operator for the qudits $S$ after tracing out the rest. Note that $Q_m$ reduces to 
the original Meyer-Wallach measure when $m=1$ and $D=2$. The above ``multipartite'' entanglement 
measures are merely averages over the well-established bipartite measure. Consequently, $Q_m$ is 
invariant under local unitary transformations, nonincreasing on average under local quantum operations 
and classical communication i.e. $Q_m$ is an entanglement monotone \cite{vidal}, and $0\leq Q_m\leq 1$. The lower bound 
is only reached for product states. 
\\

\noindent{\bf Proposition 1:} $\quad Q_m(\psi)=0 \quad\text{iff}\quad 
\ket{\psi}=\bigotimes_{j=1}^{n}\ket{\psi_j}\,$ for some $\,\ket{\psi_j}\in\mathbb{C}^D\quad$ i.e. $\ket{\psi}$ is a product state.
\\

When $m=1$ the upper bound is reached by the generalized GHZ states
\begin{equation}
\ket{\gamma}=\frac{1}{\sqrt{D}}\sum_{j=0}^{D-1}\ket{j}^{\otimes n}.
\end{equation}
In general
\begin{equation}
Q_m(\gamma)=1-\frac{D^{m-1}-1}{D^m-1}.
\end{equation}
and hence, the entangled states $\ket{\gamma}$ do not saturate the upper bound for $m>1$.
We have not, however, established whether or not there even {\em exist} states which saturate the 
upper bound. 

Define an {\em $m$-uniform} multi-qudit state to be a state with the property that 
after tracing out all but $m$ qudits we are left with the maximally mixed state, for any 
$m$-tuple of qudits. Thus, all information about the system is lost upon the removal of $n-m$ 
or more parties. 
\\

\noindent{\bf Proposition 2:} $\quad Q_m(\psi)=1 \quad\text{iff}\quad \rho_S = \tr_{S'}\ket{\psi}\bra{\psi}=D^{-m}\hat{1}\,$ whenever $\,|S|=m \quad$ 
i.e. $\ket{\psi}$ is $m$-uniform.
\\

\noindent Obviously, if $\ket{\psi}$ is $m$-uniform then it is also $(m-1)$-uniform, and hence, 
$Q_m(\psi)=1 \implies Q_{m-1}(\psi)=1$. However, note that the measures $Q_m$ do not obey any 
ordering. For example, in the case of qubits, consider the generalized 
W-states
\begin{equation}
\ket{\omega} = \frac{1}{\sqrt{n}}\sum_{j=1}^n \ket{0}^{\otimes j-1}\otimes\ket{1}\otimes\ket{0}^{\otimes n-j}.
\end{equation} 
One can calculate
\begin{equation}
Q_m(\omega)=\frac{2^{m+1}}{2^m-1}\frac{(n-m)m}{n^2}
\end{equation}
and hence, for $n=6$ say, $Q_1=5/9<Q_3=4/7<Q_2=16/27$.
The measures $Q_m$ also do not preserve the partial ordering of entangled states 
i.e. $Q_{m'}(\psi)\leq Q_{m'}(\phi)$ does not necessarily imply that $Q_m(\psi)\leq Q_m(\phi)$ for 
other $m$. These facts might be considered as unlucky properties of $Q_m$. However they do suggest 
that the extremal entanglement measure $Q_{\lfloor n/2\rfloor}$ does not necessarily tell the entire 
story; different $Q_m$ capture different aspects of multipartite entanglement. The original Meyer-Wallach 
measure $Q_1$ is the average entanglement between individual qudits and the rest, whereas, on increasing 
$m$, $Q_m$ measures the average entanglement between blocks of qudits, of an increasing size, and the rest. 
Consequently, as $m$ increases, we expect that $Q_m$ will be sensitive to correlations of an increasingly 
global nature. 

Proposition 2 implies that the task of finding states which saturate the the upper bound 1 of 
$Q_m$ is equivalent to the construction of $m$-uniform multi-qudit states. We now show in the 
next section how quantum-error-correcting codes (QECC's) produce $m$-uniform multi-qudit states. 
An example is the six-qubit hexacode state $\ket{H}$, which arises as the code subspace of 
the self-dual qubit stabilizer code $[[6,0,4]]$. In this case $Q_1(H)=Q_2(H)=Q_3(H)=1$. 

\section{Multipartite entanglement and QECC's}
\label{sec3}

The idea behind quantum error correction \cite{calderbank,gottesman,knill,preskill,nielsen,grassl,klappenecker2} is to encode quantum states
into qudits in such a way that a small number of errors affecting the individual 
qudits can be measured and corrected to perfectly restore the original encoded state.  
The encoding of a $K$-dimensional quantum state into $n$ qudits is simply a linear map 
from $\mathbb{C}^K$ to a subspace $\mathcal{Q}$ of $(\mathbb{C}^D)^{\otimes n}$. The subspace itself
is referred to as the {\it code} and is orientated in such a way 
that errors on the qudits move encoded states in a direction perpendicular to the code.

\subsection{General QECC's}

An {\it error operator} $E$ is a linear operator acting on $(\mathbb{C}^D)^{\otimes n}$. The error 
is said to be {\it detectable} by the quantum code $\mathcal{Q}$ if 
\begin{equation}
\bra{\psi}E\ket{\psi}=\bra{\phi}E\ket{\phi}
\end{equation}
for all normalized $\ket{\psi},\ket{\phi}\in\mathcal{Q}$. Equivalently, if $\mathcal{Q}$ is 
spanned by an orthonormal {\it logical basis} $\{\ket{j_L}\;|\;j=0,\dots,K-1\}$, then an error 
$E$ is detectable if and only if
\begin{equation}
\bra{j_L}E\ket{i_L}=C(E)\delta_{ij} \label{detect}
\end{equation}
for all $0\leq i,j\leq K-1$ where the constant $C(E)$ depends only on $E$. It is a general theorem of QECC's that a set of errors $\mathcal{E}$ 
can be {\it corrected} by a code $\mathcal{Q}$, if and only if for each $E_1,E_2\in\mathcal{E}$,
the error $E_2^\dag E_1$ is detectable by $\mathcal{Q}$. 

A {\it local error operator} has the form
\begin{equation}
E=M_1\otimes\dots\otimes M_n
\end{equation}
where each $M_i$ acts on $\mathbb{C}^D$. The {\it weight} of a local error operator $E$, denoted
by $\wt(E)$, is the number of elements $M_i$ which are not scalar multiples of the identity. 
A quantum code $\mathcal{Q}$ has a {\it minimum distance} of at least $d$ if and only if all local 
error operators of weight less than $d$ are detectable by $\mathcal{Q}$. A code with minimum 
distance $d=2t+1$ allows the correction of arbitrary errors affecting up to $t$ qudits. 
In the case of qubits, such codes are denoted by the triple $((n,K,d))$. We will use 
the notation $((n,K,d))_D$ for the general case of qudits \cite{rains2}. An $((n,K,d))_D$ code is 
called {\it pure} if $\bra{\psi}E\ket{\psi}=D^{-n}\tr{E}$ for all $\ket{\psi}\in\mathcal{Q}$ 
whenever $\wt(E)<d$. When considering self-dual codes ($K=1$), we adopt the convention that  
the notation $((n,1,d))_D$ refers only to pure codes since the condition on the minimum 
distance is otherwise trivial.

There is a continuum of possible errors in a single qudit; however, due to the phenomenon of 
measurement collapse, the correction of an arbitrary single-qudit error only requires an ability to correct $D^2$ 
different types, each corresponding to an orthonormal basis element for single-qudit operations. One choice 
for a {\it nice error basis} \cite{klappenecker,knill2,knill3} is the {\it displacement operator basis}  
\begin{equation}
D(\mu,\nu) \equiv e^{i\pi\mu\nu/D}X^\mu Z^\nu  \qquad  0\leq\mu,\nu\leq D-1
\end{equation}
where the Weyl operators $X$ and $Z$ are defined on a basis $\{\ket{j}\;|\;j=0,\dots,D-1\}$ for $\mathbb{C}^D$ through the equations
\begin{equation}
X\ket{j}=\ket{j+1 \text{ mod } D}, \qquad Z\ket{j}= e^{2\pi ij/D}\ket{j}.
\end{equation}
The displacement operators reduce to the Pauli matrices for qubits, satisfy the relations 
\begin{eqnarray}
D(\mu,\nu) &=& e^{i\pi\nu}D(\mu+D,\nu) \;=\; e^{i\pi\mu}D(\mu,\nu+D) \\
D(\mu,\nu)^\dag &=& D(-\mu,-\nu) \;=\; e^{i\pi(\mu+\nu+D)}D(D-\mu,D-\nu) \label{daggerD}\\
D(\mu,\nu)D(\alpha,\beta) &=& e^{2\pi i(\nu\alpha-\mu\beta)/D}D(\alpha,\beta)D(\mu,\nu) \;=\; e^{\pi i(\nu\alpha-\mu\beta)/D}D(\mu+\alpha,\nu+\beta) \\
\tr\left[D(\mu,\nu)^{\dag}D(\alpha,\beta)\right] &=& D\delta_{\mu\alpha}\delta_{\nu\beta},
\end{eqnarray}
and thus form an orthonormal basis for all single-qudit operators:
\begin{equation}
A = \frac{1}{D}\sum_{\mu,\nu=0}^{D-1}\tr\left[D(\mu,\nu)^{\dag}A\right]D(\mu,\nu).
\end{equation}
Similarly, the operators 
\begin{equation}
\mathcal{D}(\bm{\mu},\bm{\nu})\equiv\mathcal{D}(\mu_1\dots\mu_n,\nu_1\dots\nu_n)\equiv D(\mu_1,\nu_1)\otimes\dots\otimes D(\mu_n,\nu_n) \qquad 0\leq\mu_k,\nu_k\leq D-1
\end{equation}
form an orthonormal basis for the set of all $n$-qudit operators: 
$A=D^{-n}\sum_{\bm{\mu},\bm{\nu}}\tr[\mathcal{D}(\bm{\mu},\bm{\nu})^\dag A]\mathcal{D}(\bm{\mu},\bm{\nu})$. 
The weight of $\mathcal{D}(\bm{\mu},\bm{\nu})$ is simply the number of pairs $(\mu_k,\nu_k)$ 
different from $(0,0)$. We are now in a position to make a more explicit definition of what we mean by 
an $((n,K,d))_D$ QECC.  
\\

\noindent{\bf Definition:} Let $\mathcal{Q}$ be a $K$-dimensional subspace of $(\mathbb{C}^D)^{\otimes n}$ 
spanned by the orthonormal logical basis $\{\ket{j_L}\;|\;j=0,\dots,K-1\}$. Then $\mathcal{Q}$ is 
called an {\it $((n,K,d))_D$ quantum-error-correcting code} if 
\begin{equation}
\bra{j_L}\mathcal{D}(\bm{\mu},\bm{\nu})\ket{i_L}=C(\bm{\mu},\bm{\nu})\delta_{ij}
\end{equation}
for all $\mathcal{D}(\bm{\mu},\bm{\nu})$ with $\wt[\mathcal{D}(\bm{\mu},\bm{\nu})]<d$ and $0\leq i,j\leq K-1$. If 
$C(\bm{\mu},\bm{\nu})=\delta_{\bm{\mu 0}}\delta_{\bm{\nu 0}}$ the code is called {\it pure}. An 
$((n,1,d))_D$ code must be pure by convention.\\

An $((n,K,d))_D$ QECC can detect and recover all errors acting on $<d/2$ qudits. It is now evident how quantum
codes produce maximally entangled states.
\\

\noindent{\bf Proposition 3:} $\quad Q_m(\psi)=1 \quad\text{iff}\quad \ket{\psi}$ is 
a (pure) $((n,1,m+1))_D$ quantum-error-correcting code.
\\

\noindent{\bf Proof:} If $Q_m(\psi)=1$ then $\ket{\psi}$ is $m$-uniform, and consequently
\begin{eqnarray}
\bra{\psi}\mathcal{D}(\bm{\mu},\bm{\nu})\ket{\psi} &=& \tr\big[\ket{\psi}\bra{\psi}\mathcal{D}(\bm{\mu},\bm{\nu})\big] \\
&=& D^{-n}\tr\left[\mathcal{D}(\bm{\mu},\bm{\nu})\right] \:\quad(\text{whenever }\; \wt[\mathcal{D}(\bm{\mu},\bm{\nu})]\leq m)\\
&=& \delta_{\bm{\mu 0}}\delta_{\bm{\nu 0}}
\end{eqnarray}
given that the displacement operators are traceless for all $(\mu,\nu)\neq (0,0)$. Thus, $\ket{\psi}$ is an 
$((n,1,m+1))_D$ QECC. 

Conversely, if $\ket{\psi}$ is an $((n,1,m+1))_D$ QECC, then rewriting $\ket{\psi}\bra{\psi}$ in 
the displacement operator basis 
\begin{equation}
D^n\ket{\psi}\bra{\psi}\;=\;\hat{1}\quad+\sum_{1\leq\wt[\mathcal{D}(\bm{\mu},\bm{\nu})]\leq m}c_{\bm{\mu}\bm{\nu}}\mathcal{D}(\bm{\mu},\bm{\nu})\quad+\sum_{m+1\leq\wt[\mathcal{D}(\bm{\mu},\bm{\nu})]\leq n}c_{\bm{\mu}\bm{\nu}}\mathcal{D}(\bm{\mu},\bm{\nu})
\end{equation}
we see that the coefficients $c_{\bm{\mu}\bm{\nu}}=\bra{\psi}\mathcal{D}(\bm{\mu},\bm{\nu})\ket{\psi}$ are 
nonzero only in the second sum, and hence, given the traceless property of the displacement operators,
\begin{equation}
\rho_S = \tr_{S'}\ket{\psi}\bra{\psi}= D^{-m}\hat{1}
\end{equation}
whenever $|S|=m$. Thus $\ket{\psi}$ is $m$-uniform and $Q_m(\psi)=1$. $\Box$
\\

Note that any state $\ket{\psi}\in\mathcal{Q}$, where $\mathcal{Q}$ is a pure $((n,K,m+1))_D$ QECC, 
is itself an $((n,1,m+1))_D$ QECC. Consequently, pure $((n,K,m+1))_D$ codes define entire subspaces of maximally 
entangled states. The connection between quantum codes and entanglement is noted in 
\cite{preskill}, and alluded to elsewhere \cite{knill,bravyi}; however, we cite the work of Rains 
\cite{rains4} for a rigorous proof of the relationship even though no mention of entanglement can be
found in the paper. Here quantum weight enumerators were studied extensively. It will later prove  
advantageous to now revisit Rains' work in the current article.

Defining $P_\mathcal{Q}$ as the projector onto the code subspace $\mathcal{Q}$ with dimension $K$, 
the Shor-Laflamme enumerators of a quantum code are \cite{shor}
\begin{eqnarray}
A_i(P_\mathcal{Q})&=&\frac{1}{K^2}\sum_{\wt\left[\mathcal{D}(\bm{\mu},\bm{\nu})\right]=i}\big|\tr[\mathcal{D}(\bm{\mu},\bm{\nu})P_\mathcal{Q}]\big|^2 \\
B_i(P_\mathcal{Q})&=&\frac{1}{K}\sum_{\wt\left[\mathcal{D}(\bm{\mu},\bm{\nu})\right]=i}\tr[\mathcal{D}(\bm{\mu},\bm{\nu})P_\mathcal{Q}\mathcal{D}(\bm{\mu},\bm{\nu})^\dag P_\mathcal{Q}]
\end{eqnarray}
where $i=0,\dots,n$. Rains \cite{rains4} defined two new enumerators
\begin{eqnarray}
A_i'(P_\mathcal{Q})&=&\frac{1}{K^2}\sum_{|S|=i}\tr_S\big[\tr_{S'}[P_\mathcal{Q}]^2\big] \\
B_i'(P_\mathcal{Q})&=&\frac{1}{K}\sum_{|S|=i}\tr_{S'}\big[\tr_S[P_\mathcal{Q}]^2\big]
\end{eqnarray}
related to the Shor-Laflamme enumerators via the equations
\begin{eqnarray}
A_m'(P_\mathcal{Q})&=& D^{-m} \sum_{i=0}^m \frac{(n-i)!}{(m-i)!(n-m)!}A_i(P_\mathcal{Q}) \\
B_m'(P_\mathcal{Q})&=& D^{-m} \sum_{i=0}^m \frac{(n-i)!}{(m-i)!(n-m)!}B_i(P_\mathcal{Q}).
\end{eqnarray}
This relationship was only given in the qubit case where the displacement operators reduce 
to Hermitian Pauli matrices. However the proof extends easily to qudits with the the help of 
Eq. (\ref{daggerD}). It is easy to see that the weight enumerators satisfy the normalization condition
$A_0'(P_\mathcal{Q})=B_0'(P_\mathcal{Q})=A_0(P_\mathcal{Q})=B_0(P_\mathcal{Q})=1$, and for self-dual codes ($K=1$) 
\begin{equation}
B_i'(P_\mathcal{Q})=A_i'(P_\mathcal{Q}) \qquad \Big[\,B_i(P_\mathcal{Q})=A_i(P_\mathcal{Q})\,\Big] 
\end{equation}
for all $0\leq i\leq n$. In general, the weight enumerators satisfy \cite{rains4}
\begin{equation}
B_i'(P_\mathcal{Q})\geq A_i'(P_\mathcal{Q})\geq 0 \qquad \Big[\,B_i(P_\mathcal{Q})\geq A_i(P_\mathcal{Q})\geq 0\,\Big]
\end{equation}
for all $0\leq i\leq n$. 
\\

\noindent{\bf Theorem \cite{rains4}:} Let $\mathcal{Q}$ be a quantum code with associated 
projector $P_\mathcal{Q}$. Then $\mathcal{Q}$ has minimum distance of at least $d$ iff 
\begin{equation}
B_{d-1}'(P_\mathcal{Q})=A_{d-1}'(P_\mathcal{Q}) \qquad \Big[\,B_i(P_\mathcal{Q})=A_i(P_\mathcal{Q}) \,\text{ for all }\, 0<i<d \,\Big]
\end{equation}
and is pure iff 
\begin{equation}
B_{d-1}'(P_\mathcal{Q})=A_{d-1}'(P_\mathcal{Q})=\frac{D^{1-d}n!}{(d-1)!(n-d+1)!} \qquad \Big[\,B_i(P_\mathcal{Q})=A_i(P_\mathcal{Q})=0 \,\text{ for all }\, 0<i<d \,\Big].
\end{equation}
\\

Proposition 3 is now immediately apparent since 
\begin{equation}
Q_m(\psi)=\frac{D^m}{D^m-1}\bigg[1-\frac{m!(n-m)!}{n!}A_m'\big(\ket{\psi}\bra{\psi}\big)\bigg]=1-\frac{1}{D^m-1}\sum_{i=1}^m \frac{m!(n-i)!}{n!(m-i)!}A_i\big(\ket{\psi}\bra{\psi}\big)\label{Wt2Qm}. 
\end{equation}
Noting that $KA'_i=B_{n-i}'$, one can use the above theorem to derive bounds on the minimum distance 
for general quantum codes. In the case of $((n,1,d))_D$ QECC's the following conditions must hold:
\begin{eqnarray}
A'_i &=& A_{n-i}'  \qquad\,\, 0\leq i\leq n\label{lp1} \\  A_0 &=& 1  \label{lp2} \\  
A_i &=& 0 \qquad\qquad 0<i<d \label{lp3} \\ A_i &\geq& 0 \qquad\qquad d\leq i\leq n .\label{lp4}
\end{eqnarray}
When $d=\lfloor n/2\rfloor+1$, these equations uniquely specify the weight distribution $\{A_i\}$. 
Solving equations (\ref{lp1})-(\ref{lp3}), we obtain
\begin{equation}
A_i=\frac{n!}{(n-i)!}\sum_{j=d}^i\frac{(-1)^{i-j}(D^{2j-n}-1)}{j!(i-j)!} \qquad d\leq i\leq n
\label{MDSweightdist}\end{equation}
and under the condition $A_{d+1}\geq 0$, we find that we at least require
\begin{equation}
n \leq \left\{\begin{array}{ll} 
  2(D^2-1) & \text{ if } n \text{ is even } \\
 2D(D+1)-1 & \text{ if } n \text{ is odd } 
\end{array}\right. \label{lpbound1}
\end{equation}
for an $((n,1,\lfloor n/2\rfloor+1))_D$ QECC to exist. Consequently, $Q_{\lfloor n/2\rfloor}(\psi)<1$ 
for all $\ket{\psi}$ whenever Eq. (\ref{lpbound1}) is not satisfied. For $d\leq\lfloor n/2\rfloor$ 
we must resort to linear programming techniques on equations (\ref{lp1})-(\ref{lp4}) to prove the 
nonexistence of $((n,1,d))_D$ QECC's. However tighter bounds could be obtained by using the 
generalized quantum shadow enumerators \cite{rains5,rains6}. We make no attempt at this task in the 
current article, but instead specialize to qubits where many examples of QECC's are already known.

\subsection{Stabilizer qubit QECC's}

An important class of quantum codes are the so-called {\it additive} or {\it stabilizer} codes 
\cite{calderbank,gottesman}. A {\it stabilizer code} is defined as a joint eigenspace 
of an Abelian subgroup $S$ (called the {\it stabilizer}) of the {\it error group} 
$\mathcal{E}=\{\pm e^{i\pi\lambda/D}\mathcal{D}(\bm{\mu},\bm{\nu})\,|\,0\leq\mu_k,\nu_k,\lambda\leq D-1\}$.
When $D$ is prime, these codes can be described by an $(n-k)\times n$ stabilizer matrix over 
$GF(D^2)$ and are examples of $((n,D^k,d))_D$ QECC's. The notation $[[n,k,d]]_D$ is then used, 
or simply $[[n,k,d]]$ when $D=2$. 

A classical {\it additive code over $GF(4)$ of length $n$} is an additive subgroup $\mathcal{C}$ of $GF(4)^n$.
In the case of qubits, stabilizer codes correspond to classical additive codes over $GF(4)$ \cite{calderbank}. 
This is shown as follows. Letting $GF(4)=\{0,1,\w,\wb\}$ where 
$\wb=\w^2=1+\w$, we define the {\it conjugate} of $x\in GF(4)$, denoted $\overline{x}$, by the mapping 
$\overline{0}=0$, $\overline{1}=1$, and $\overline{\wb}=\w$. Next define the {\it trace} map 
$\Tr:GF(4)\rightarrow GF(2)$ by $\Tr(x)=x+x^2$ i.e. $\Tr(0)=\Tr(1)=0$ and $\Tr(\w)=\Tr(\wb)=1$, 
and the {\it trace inner product} of two vectors ${\bf x} = x_1\dots x_n$ and ${\bf y} = y_1\dots y_n$ in $GF(4)^n$
as
\begin{equation}
{\bf x\star y} = \sum_{i=1}^n\Tr\big(x_i\overline{y_i}\big) \;\;\in GF(2).
\end{equation} 
The {\it weight} $\wt({\bf x})$ of ${\bf x}\in GF(4)^n$ is the number of nonzero 
components of ${\bf x}$, and the {\it minimum weight} of a code $\mathcal{C}$ is the smallest 
weight of any nonzero codeword in $\mathcal{C}$. Next, by defining the mapping $\Phi:GF(4)^n\rightarrow\mathcal{E}$ by $\Phi({\bf x})=\mathcal{D}(\phi^{-1}({\bf x}))$ where 
$\phi(\bm{\mu},\bm{\nu})=\w\bm{\mu}+\wb\bm{\nu}$, we can associate elements of $GF(4)$ with Pauli matrices 
($\w\rightarrow X$, $\wb\rightarrow Z$, $1\rightarrow iXZ$, $0\rightarrow I$), 
addition of vectors over $GF(4)^n$ with multiplication of operators in $\mathcal{E}$ 
(neglecting phases), and the trace inner product on $GF(4)^n$ with the commutator on 
$\mathcal{E}$. 

If $\mathcal{C}$ is an additive code, its {\it dual} is the additive code 
$\mathcal{C}^\perp=\{{\bf x}\in GF(4)^n \,|\, {\bf x\star c}=0 \;\forall\, {\bf c}\in\mathcal{C}\}$.
The code $\mathcal{C}$ is called {\it self-orthogonal} if $\mathcal{C}\subseteq\mathcal{C}^\perp$ and 
{\it self-dual} if $\mathcal{C}=\mathcal{C}^\perp$. The following theorem now applies \cite{calderbank}:
Suppose $\mathcal{C}$ is a self-orthogonal additive subgroup of $GF(4)^n$, containing $2^{n-k}$ vectors, such that 
there are no vectors of weight $<d$ in $\mathcal{C}^\perp\backslash\mathcal{C}$. Then any joint eigenspace of 
$\Phi(\mathcal{C})$ is an $[[n,k,d]]$ QECC. 

We say that $\mathcal{C}$ is {\it pure} if there are no nonzero vectors of weight $<d$ in $\mathcal{C}^\perp$. 
The associated QECC is then pure if and only if $\mathcal{C}$ is pure. By convention, an $[[n,0,d]]$ QECC
corresponds to a self-dual additive code $\mathcal{C}$ with minimum weight $d$. Consequently, 
$[[n,0,d]]$ QECC's are always pure, and are examples of $((n,1,d))$ QECC's which saturate the 
entanglement measures $Q_m$. 

The advantage of making the above correspondence is that a wealth of classical 
coding theory immediately becomes available. Indeed the classical self-dual additive {\it hexacode} with 
generator matrix
\begin{equation}
\left[\begin{array}{cccccc} 1&0&0&1&\w&\w \\ 0&1&0&\w&1&\w \\ 0&0&1&\w&\w&1 \\ \w&0&0&\w&\wb&\wb \\ 0&\w&0&\wb&\w&\wb \\ 0&0&\w&\wb&\wb&\w
\end{array}\right] \label{genhex}
\end{equation}
gives the quantum hexacode $[[6,0,4]]$ mentioned previously. The rows of the generator matrix define 
a basis (under addition) for the classical code $\mathcal{C}$, and, with the above correspondence, define generators 
(up to a phase) for the stabilizer $S$ in the quantum version. Another example is the $[[2,0,2]]$ qubit code generated by
\begin{equation}
\left[\begin{array}{cc} 1&1 \\ \w&\w 
\end{array}\right].
\end{equation}
In this case the quantum code is an EPR state e.g. $(\ket{00}+\ket{11})/\sqrt{2}$. We can obtain an 
$[[5,0,3]]$ qubit code by deleting the first row and column of the hexacode generator matrix [Eq. (\ref{genhex})]. 
This process is called {\it shortening} \cite{gaborit}. A $[[3,0,2]]$ code 
\begin{equation}
\left[\begin{array}{ccc} 1&1&0 \\ \w&\w&\w \\ 1&0&1 
\end{array}\right]
\end{equation}
is obtained by {\it lengthening} the $[[2,0,2]]$ code. 

The four codes of lengths $n=2,3,5$ and $6$ mentioned thus far all produce quantum stabilizer 
codes with the property $Q_{\lfloor n/2\rfloor}(\psi)=1$. Unfortunately, known bounds on such 
codes prevent this from being the case for other lengths. An additive self-dual code is called 
{\it Type II} if all codewords have even weight, and {\it Type I} otherwise. It can be shown 
that all Type II codes have even length. If $d_I$, $d_{II}$ is the minimum weight of an 
additive self-dual Type I, Type II code, respectively, of length $n>1$, then \cite{rains,rains3,gaborit} 
\begin{eqnarray}
d_I & \leq & \left\{\begin{array}{ll} 
 2\lfloor n/6 \rfloor +1 & \text{ if } n \equiv 0 \text{ mod } 6 \\
 2\lfloor n/6 \rfloor +3 & \text{ if } n \equiv 5 \text{ mod } 6 \\
 2\lfloor n/6 \rfloor +2 & \text{ otherwise }
\end{array}\right. \label{rainsbound}\\
d_{II} & \leq & 2\lfloor n/6 \rfloor +2. \label{rainsbound2}
\end{eqnarray}
If a code meets the appropriate bound it is called {\it extremal}. A code is called {\it optimal} 
when it is not extremal and no code can exist with a larger minimum weight. The above bounds 
imply that $[[n,0,\lfloor n/2\rfloor+1]]$ stabilizer codes may exist only when $n=2,3,5,6$ and 7.
However $[[7,0,3]]$ codes are known to be optimal and we a left with the remaining four cases. 

The {\it weight distribution} of an additive code $\mathcal{C}$
\begin{equation}
A_i \equiv |\{{\bf x}\in\mathcal{C}\,|\,\wt({\bf x})=i\}|
\end{equation}
is also the weight distribution for the corresponding quantum stabilizer code,
and thus, the entanglement of the stabilized state is easily calculated through formula 
(\ref{Wt2Qm}). We can see this by noting that for stabilizer codes, the projection 
onto $\mathcal{Q}$ is given by \cite{klappenecker2}
\begin{equation}
P_\mathcal{Q}=\frac{1}{|S|}\sum_{E\in S}\lambda(E)^{-1}E
\end{equation}
where $S$ is the stabilizer and $\lambda(E)$ is the eigenvalue associated with $E$ 
i.e. $E\ket{\psi}=\lambda(E)\ket{\psi}$ for all $\ket{\psi}\in\mathcal{Q}$. We remark that the 
quantum weight distribution $B_i$ corresponds to the classical weight distribution 
of the dual code $\mathcal{C^\perp}$.

\begin{table}
\begin{tabular}{r|r|rrrrrrrrrrrrrr|cccccc}
$n$ & $d$ & $A_0$ & $A_1$ & $A_2$ & $A_3$ & $A_4$ & $A_5$ & $A_6$ & $A_7$ & $A_8$ & $A_9$ & $A_{10}$ & $A_{11}$ & $A_{12}$ & $A_{13}$ & $Q_1$ & $Q_2$ & $Q_3$ & $Q_4$ & $Q_5$ & $Q_6$\\
\hline\hline
2 & 2 & 1 & 0 & 3 &&&&&&&&&&&& 1 & \\
\hline
3 & 2 & 1 & 0 & 3 & 4 &&&&&&&&&&& 1 & \\
\hline
4 & 2 & 1 & 0 & 6 & 0 & 9 &&&&&&&&&& 1 & 2/3 \\
4 & 2 & 1 & 0 & 2 & 8 & 5 &&&&&&&&&& 1 & 8/9 \\
\hline
5 & 3 & 1 & 0 & 0 & 10 & 15 & 6 &&&&&&&&& 1 & 1 \\
\hline
6 & 4 & 1 & 0 & 0 & 0 & 45 & 0 & 18 &&&&&&&& 1 & 1 & 1 \\
\hline
7 & 3 & 1 & 0 & 0 & 7 & 21 & 42 & 42 & 15 &&&&&&& 1 & 1 & 34/35 \\
7 & 3 & 1 & 0 & 0 & 3 & 29 & 42 & 34 & 19 &&&&&&& 1 & 1 & 242/245 \\
\hline
8 & 4 & 1 & 0 & 0 & 0 & 42 & 0 & 168 & 0 & 45 &&&&&& 1 & 1 & 1 & 24/25 \\
8 & 4 & 1 & 0 & 0 & 0 & 26 & 64 & 72 & 64 & 29 &&&&&& 1 & 1 & 1 & 512/525 \\
\hline
9 & 4 & 1 & 0 & 0 & 0 & 26 & 48 & 136 & 160 & 93 & 48 &&&&& 1 & 1 & 1 & 932/945 \\
9 & 4 & 1 & 0 & 0 & 0 & 18 & 72 & 120 & 144 & 117 & 40 &&&&& 1 & 1 & 1 & 104/105 \\
\hline
10 & 4 & 1 & 0 & 0 & 0 & 30 & 0 & 300 & 0 & 585 & 0 & 108 &&&& 1 & 1 & 1 & 104/105 & 212/217 \\
\hline
11 & 5 & 1 & 0 & 0 & 0 & 0 & 66 & 198 & 330 & 495 & 550 & 330 & 78 &&& 1 & 1 & 1 & 1 & 216/217 \\
\hline
12 & 6 & 1 & 0 & 0 & 0 & 0 & 0 & 396 & 0 & 1485 & 0 & 1980 & 0 & 234 && 1 & 1 & 1 & 1 & 1 & 146/147 \\
\hline
13 & 5 & 1 & 0 & 0 & 0 & 0 & 15 & 236 & 356 & 1197 & 1530 & 2012 & 1956 & 650 & 239 & 1 & 1 & 1 & 1 & 13294/13299 & 26938/27027 \\

\end{tabular}
\caption{\label{tab} The weight distributions $A_i$ and corresponding entanglement $Q_m$ for extremal 
(or optimal for $n=7$ and 13) additive self-dual codes. In all but the cases $n=10$ and 13 
these are the only possible weight distributions.}
\end{table}

In Table \ref{tab} the weight distributions for extremal (or optimal for $n=7$ and 13) additive 
self-dual codes is collected \cite{hohn,gaborit2}. In all but the cases $n=10$ and 13 these are the 
only possible weight distributions. The weight distribution is unique for extremal Type II codes.  
Also tabulated is the corresponding entanglement $Q_m$ for the quantum code.

Although the bounds mentioned above [Eq.'s (\ref{rainsbound},\ref{rainsbound2})] were given in the context 
of stabilizer codes, they also apply to general QECC's \cite{rains5}. Consequently, for qubits, there 
exist states $\ket{\psi}$ with $Q_{\lfloor n/2\rfloor}(\psi)=1$ only in the cases $n=2,3,5,6$, and possibly 
when $n=7$ where a nonadditive $((7,1,4))$ code might still exist. It is not 
known what the supremum of $Q_m(\psi)$ is in general; however, given the examples in Table \ref{tab}, we expect
it to be very close to 1 when $n$ is large. It is interesting that the mean of $Q_m(\psi)$ (given in the next 
section) seems unaffected by the erratic behavior in the supremum. 

For the most part, quantum coding theorists have primarily studied qubit codes. Some work on qudit 
codes exists \cite{rains2,ashikhmin,schlingemann,feng,grassl2}, but there are very few known examples. 
One exception is the generalization of the hexacode. A $((6,1,4))_D$ code is known to exist for all 
$D$ \cite{rains2}. The $((6,1,4))_D$ code belongs to a class of optimal codes called maximum distance 
separable (MDS) codes which saturate the quantum Singleton bound \cite{rains2} i.e. $K=D^{n-2d+2}$. 
Quantum MDS codes must be pure, and thus, define subspaces of maximally entangled states. 
Self-dual quantum MDS codes have weight distributions specified by Eq. (\ref{MDSweightdist}).
Other examples of MDS codes include the $[[6,2,3]]_D$ and $[[7,3,3]]_D$ stabilizer codes which 
exist for all prime $D$ \cite{feng}. More recently, the existence of some families of quantum MDS codes 
was proven \cite{grassl2}. For example, when $D$ is a prime power and $n$ is even, a self-dual 
$((n,1,n/2+1))_D$ code exists for all $3\leq n\leq D$.  

\section{Multipartite entanglement over subspaces}
\label{sec4}

Using Lubkin's formula \cite{lubkin} for the average subsystem purity, one can easily 
calculate the mean entanglement for random pure states sampled according 
to the unitarily invariant Haar measure $d\mu$ ($\int d\mu(\psi)=1$):
\begin{equation}
\big\langle Q_m(\psi) \big\rangle_{\psi} \equiv \int d\mu(\psi)Q_m(\psi) = 1 - \frac{D^m+1}{D^n+1}.
\label{random}\end{equation}  
This shows that when the overall dimension $D^n$ is large, a typical state has nearly 
maximal entanglement.
One could also consider the average entanglement over a subspace $\mathcal{V}$ determined by the 
projector $P_\mathcal{V}$:
\begin{equation}
\big\langle Q_m(\psi) \big\rangle_{\psi\in\mathcal{V}} \equiv \int_{\mathcal{V}} d\mu_{\mathcal{V}}(\psi)Q_m(\psi).
\end{equation}

\noindent{\bf Proposition 4:} 
\begin{equation}
\big\langle Q_m(\psi) \big\rangle_{\psi\in\mathcal{V}}=\frac{D^m}{D^m-1}\Bigg\{1-\frac{m!(n-m)!}{n!K(K+1)}\sum_{|S|=m}\bigg(\tr_S\big[\tr_{S'}[P_\mathcal{V}]^2\big]+\tr_{S'}\big[\tr_S[P_\mathcal{V}]^2\big]\bigg)\Bigg\} 
\end{equation}
where $K=\tr P_\mathcal{V}=\dim \mathcal{V}$.
\\

\noindent{\bf Proof:} Consider an arbitrary bipartite system 
$\mathcal{H}=\mathbb{C}^{D_A}\otimes\mathbb{C}^{D_B}$ and define the swap operators $T_{ij}$ 
($1\leq i<j\leq 4$) which transpose the $i$-th and $j$-th factors of $\mathcal{H}^{\otimes 2}$.
Using the identity $\tr [(A\otimes B) T]=\tr [AB]$, where $T$ is the swap, we first rewrite the 
subsystem purity of a state $\ket{\psi}\in\mathcal{H}$ as 
\begin{equation}
\tr{\rho_A}^2 = \tr\big[ \ket{\psi}\bra{\psi}^{\otimes 2} T_{13}\big]
\end{equation}
where $\rho_A=\tr_B\ket{\psi}\bra{\psi}$. Now consider the operator 
\begin{equation}
\omega\equiv\int_\mathcal{V} d\mu_{\mathcal{V}}(\psi)\ket{\psi}\bra{\psi}^{\otimes 2} 
\end{equation}
supported on the totally symmetric subspace $P\mathcal{V}^{\otimes 2}$, 
where the projector $P=(1+T_{13}T_{24})/2$. If we choose $d\mu_\mathcal{V}$ to be the 
unitarily invariant Haar measure on $\mathcal{V}$, then $[U^{\otimes 2},\omega]=0$ for all 
unitary operators $U\in\text{U}(D_AD_B)$. And since the group elements $U^{\otimes 2}$ act 
irreducibly on $P\mathcal{V}^{\otimes 2}$, by Schur's lemma \cite{tung}, $\omega$ is simply 
a scalar multiple of the identity (on $P\mathcal{V}^{\otimes 2}$). Hence,
\begin{equation} 
\omega=\frac{2}{K(K+1)}{P_\mathcal{V}}^{\otimes 2}P 
\end{equation}
on $\mathcal{H}^{\otimes 2}$, where the constant factor is found through the normalization condition $\tr\omega=1$. Thus
\begin{eqnarray}
\int_\mathcal{V} d\mu_{\mathcal{V}}(\psi)\tr{\rho_A}^2 &=& \frac{1}{K(K+1)}\tr\big[{P_\mathcal{V}}^{\otimes 2} (1+T_{13}T_{24})T_{13}\big]\\
 &=& \frac{1}{K(K+1)}\Big(\tr\big[{P_\mathcal{V}}^{\otimes 2}T_{13}\big]+\tr\big[{P_\mathcal{V}}^{\otimes 2}T_{24}\big]\Big) \\
 &=& \frac{1}{K(K+1)}\big(\tr{\mathop{\tilde{\rho}_A}}^2+\tr{\mathop{\tilde{\rho}_B}}^2\big)
\end{eqnarray}
where $\tilde{\rho}_A=\tr_BP_\mathcal{V}$ and $\tilde{\rho}_B=\tr_AP_\mathcal{V}$. We have 
derived the average purity over a subspace for an arbitrary bipartite system. Given that 
the measures $Q_m$ are simply averages over bipartite purities, 
one can now deduce the final result. $\Box$ 
\\

In particular, for a QECC $\mathcal{Q}$, one can now explicitly determine the average entanglement of 
encoded states in terms of the weight distribution of the code. For example
\begin{eqnarray}
\big\langle Q_m(\psi)\big\rangle_{\psi\in\mathcal{Q}} &=& \frac{D^m}{D^m-1}\bigg\{1-\frac{m!(n-m)!}{n!(K+1)}\Big[KA_m'\big(P_\mathcal{Q}\big)+B_m'\big(P_\mathcal{Q}\big)\Big]\bigg\} \\
&=& 1-\frac{1}{(D^m-1)(K+1)}\sum_{i=1}^m \frac{m!(n-i)!}{n!(m-i)!}\Big[KA_i\big(P_\mathcal{Q}\big)+B_i\big(P_\mathcal{Q}\big)\Big] .
\end{eqnarray} 
Such formulae make explicit the importance of entanglement as a resource for quantum 
error correction. Pure codes ($B_i=A_i=0$, $0<i<d$) necessarily have high levels of 
entanglement, but impure codes ($B_i=A_i>0$, $0<i<d$) need not. However the most 
compact codes (least $n$) all seem to be pure \cite{shor}. The relationship 
between entanglement and QECC's remains relatively unexplored in the literature.
We do not, however, pursue this line of research any further in the current 
article. Elements of the proof of Proposition 4 were borrowed from the work of 
Zanardi {\it et al.} \cite{zanardi} where the concept of entangling power was defined. In the next 
section we investigate multipartite entangling power with respect to the measures $Q_m$. 

\section{Multipartite entangling power}
\label{sec5}

Following the work of Zanardi {\it et al.} \cite{zanardi}, we define the 
{\it multipartite entangling power} of the unitary operator $U\in\text{U}(D^n)$ 
acting on $(\mathbb{C}^D)^{\otimes n}$ as simply the average entanglement generated over 
all product states:
\begin{equation}
e_p(U)\equiv\int d\mu_n(\psi_1,\dots,\psi_n)E(U\ket{\psi_1}\otimes\dots\otimes\ket{\psi_n})
\label{epower}
\end{equation}
where $\ket{\psi_i}\in\mathbb{C}^D$. The measure $d\mu_n$ is chosen to be the product 
of $n$ independent Haar measures over the constituent subsystems $\mathbb{C}^D$. 
Consequently, the entangling power is invariant under the action of local unitaries: 
$e_p(U_1\otimes\dots\otimes U_n U V_1\otimes\dots\otimes V_n)=e_p(U)$ for all $U_i, V_i\in
\text{U}(D)$. If we now restrict our attention to the entanglement measures $E=Q_m$, 
the calculation of $e_p$ is facilitated by a simple formula.
\\

\noindent{\bf Proposition 5:}  
\begin{equation}
e_p^{Q_m}(U)=\frac{D^m}{D^m-1}\Bigg(1-\frac{m!(n-m)!}{n!}\sum_{|S|=m}R_S(U)\Bigg)
\end{equation}
where the average subsystem purities
\begin{equation}
R_S(U)=\left(\frac{2}{D(D+1)}\right)^n\tr\left[ U^{\otimes 2}\,\Bigg(\prod_{i=1}^n P_{i,i+n}\Bigg)\,U^{\dag\otimes 2}\,\Bigg(\prod_{i\in S} T_{i,i+n}\Bigg)\right],
\label{avgpurity}\end{equation}
the swap operators $T_{ij}$ ($1\leq i<j\leq 2n$) transpose the $i$-th and $j$-th factors of 
$({\mathbb{C}^D})^{\otimes 2n}$, and $P_{ij}\equiv (1+T_{ij})/2$. 
\\

\noindent{\bf Proof:} The derivation of this formula is similar to that for Proposition 4 and 
follows Zanardi {\it et al.} \cite{zanardi}. We first rewrite the subsystem purity of a state 
$\ket{\Psi}\in({\mathbb{C}^D})^{\otimes n}$ as 
\begin{equation}
\tr{\rho_S}^2 = \tr\Bigg( \ket{\Psi}\bra{\Psi}^{\otimes 2}\prod_{i\in S} T_{i,i+n}\Bigg).
\end{equation}
By choosing $\ket{\Psi}=U\ket{\psi_1}\otimes\dots\otimes\ket{\psi_n}$ we have
\begin{eqnarray}
R_S(U) &\equiv& \int d\mu_n(\psi_1,\dots,\psi_n)\tr{\rho_S}^2 \\
&=& \tr\Bigg( U^{\otimes 2}\,\Omega\,U^{\dag\otimes 2}\prod_{i\in S} T_{i,i+n}\Bigg)
\end{eqnarray}
where
\begin{equation}
\Omega\equiv\int d\mu_n(\psi_1,\dots,\psi_n)\Big(\ket{\psi_1}\bra{\psi_1}\otimes\dots\otimes\ket{\psi_n}\bra{\psi_n}\Big)^{\otimes 2}.
\end{equation}
Now, considering the operator $\omega\equiv\int d\mu_1(\psi)\ket{\psi}\bra{\psi}^{\otimes 2}$ 
supported on the totally symmetric subspace $P_{12}({\mathbb{C}^D})^{\otimes 2}$, where $P_{12}=(1+T_{12})/2$,
we know from previous results (see proof of Prop. 4) that $\omega=2/D(D+1)P_{12}$ on $({\mathbb{C}^D})^{\otimes 2}$. 
Finally, given that $\Omega$ factorizes into the product of $n$ independent averages of the 
form $\omega$, we have 
\begin{equation}
\Omega=\left(\frac{2}{D(D+1)}\right)^n\prod_{i=1}^n P_{i,i+n}
\end{equation}
and our final result. $\Box$

Given our definition of the entangling power [Eq. (\ref{epower})], the (Haar measure) average of 
$e_p(U)$ over $\text{U}(D^n)$ is equivalent to the average entanglement found in random states:
\begin{equation}
\big\langle e_p^{Q_m}(U) \big\rangle_U = \big\langle Q_m(\psi) \big\rangle_\psi = 1 - \frac{D^m+1}{D^n+1}.
\label{random2}\end{equation}  
Thus, typical unitaries generate nearly maximal entanglement when the overall dimension 
$D^n$ is large.

\section{An application and conclusion} 
\label{sec6}
\begin{figure}[t]
\includegraphics[scale=1]{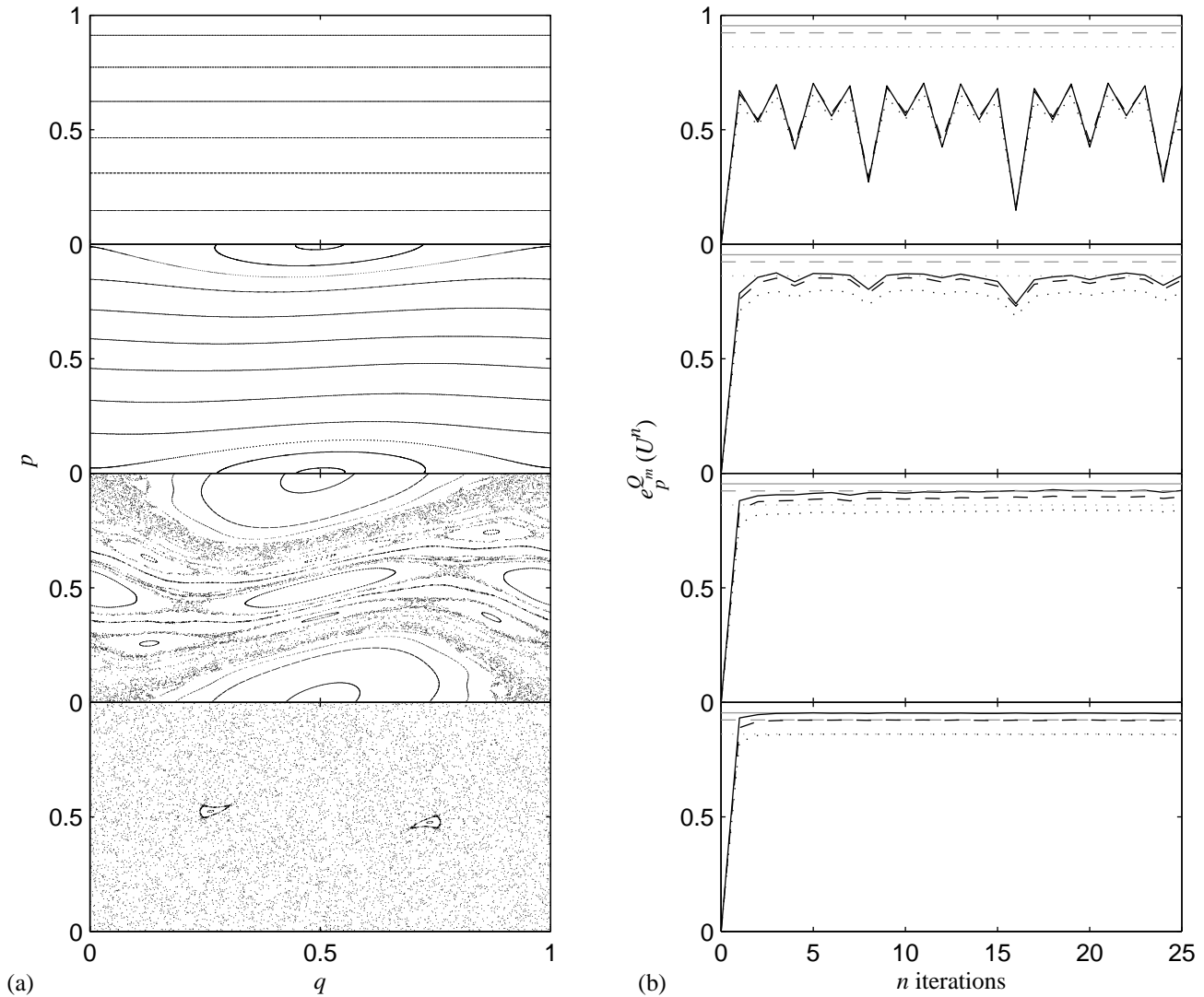}
\caption{(a) Phase-space portraits of the classical kicked rotor for $k=0,0.2,1$ and 6 
(from top to bottom), and (b) the corresponding entangling power in the quantum case. A Hilbert 
space of 6 qubits was chosen allowing the investigation of the multipartite entanglement measures 
$Q_m$ for $m=1$ (dotted), 2 (dashed) and 3 (full). The average entanglement for random states is 
shown in the lighter tones.}
\label{fig1}
\end{figure}

An immediate application of Proposition 5 (and \cite{zanardi}) occurs in the study of the entangling 
capabilities of chaotic systems \cite{sakagami,tanaka,furuya,angelo,miller,lakshminarayan,bandyopadhyay,tanaka2,fujisaki,lahiri,lakshminarayan2,bettelli,scott,bandyopadhyay2,jacquod,rossini}. 
Consider a classical map of the toroidal phase space $[0,1)^2$. A quantized version may be constructed 
in a Hilbert space of dimension $N$ spanned by the position states $\ket{q_j}$, where $q_j=(j+1/2)/N$ 
and $j=0,\dots,N-1$. By choosing $N=D^n$, we can map our Hilbert space onto the tensor-product space 
$(\mathbb{C}^D)^{\otimes n}$ through the correspondence
\begin{equation}
\ket{q_j}=\ket{x_1}\otimes\dots\otimes\ket{x_n}, \qquad\quad j=\sum_{i=1}^n x_i D^{n-i}, 
\qquad\quad x_i\in\{0,\dots,D-1\}, 
\end{equation}
and hence, use the measures $Q_m$ to investigate the quantum map's multipartite entangling power.
The different constituent qudits $x_i$ address the coarse (small $i$) and fine (large $i$) scales of 
position. Consequently, for chaotic maps where phenomena such as mixing and exponential sensitivity
are generic, we expect high levels of entanglement generation. This was noted in \cite{scott} 
where the entangling power of the quantum baker's map \cite{balazs,saraceno,schack,soklakov,tracy} was 
investigated.

For example, consider the kicked rotor (or standard map) \cite{lichtenberg}:
\begin{eqnarray}
q_{n+1} &=& q_n + p_{n+1} \mod 1\\
p_{n+1} &=& p_n +\frac{k}{2\pi}\sin 2\pi q_n \mod 1.
\end{eqnarray}
The entangling power of the quantum version \cite{hannay}
\begin{equation}
U\ket{q_j} = e^{-i \frac{kN}{2\pi}\cos 2\pi q_j} \sum_{l=0}^{N-1} e^{i\frac{\pi}{N}(l-j)^2}\ket{q_l} \qquad (N \text{ even})
\end{equation} 
constructed in a Hilbert space of 6 qubits ($N=2^6$) is plotted in Fig. \ref{fig1}(b). 
Here we choose the parameter values $k=0,0.2,1$ and 6 (from top to bottom), corresponding to the 
classical phase spaces drawn in Fig. \ref{fig1}(a). As expected, the entanglement saturates at 
a value predicted for random states [Eq. (\ref{random})] upon the appearance of chaos in the 
classical map. An alternative interpretation may be that quantized chaotic maps produce unitaries 
whose powers are typical in the space of all unitaries. This follows from Eq. (\ref{random2}).

In conclusion, we have shown that the average bipartite entanglement, $Q_m$, is a useful measure of 
multipartite entanglement, presenting a relationship between these measures and quantum-error-correcting 
codes. This was done by deriving an explicit formula relating the weight distribution of the code 
to the average entanglement of encoded states. We have also extended the work of Zanardi {\it et al.} 
\cite{zanardi} on entangling power to the multipartite case. Although the entanglement measures 
considered in this paper provide little intellectual gratification, their simplicity allows 
perhaps more important attributes such as computability and applicability. We must stress, 
however, that in defining such simple measures we offer no progress towards a deeper 
understanding of the nature of entanglement in the multipartite case.

\begin{acknowledgments}
The author would like to thank Carlton Caves and Bryan Eastin for helpful discussions. 
This work was supported in part by ONR Grant No.~N00014-00-1-0578 
and by ARO Grant No.~DAAD19-01-1-0648.
\end{acknowledgments}

\end{document}